# The Ethical Aspects of Choosing a Nuclear Fuel Cycle

Vitaly Pronskikh[1]

**Abstract:** In this paper, we addressed the problem of choosing a nuclear fuel cycle. Ethical problems related to the choice of a nuclear fuel cycle, such as the depletion of natural uranium reserves, the accumulation of nuclear waste, and the connection with the problems of nonidentity and distributive justice are considered. We examined cultural differences in attitudes toward nuclear safety and the associated ambiguities in the choice of a nuclear fuel cycle. We suggested that the reduction in consumption of natural uranium does not seem to be a feasible way of reducing nuclear waste because of the nonidentity problem.

## 1. Introduction

The issue of nuclear fuel cycles (NFCs) is closely related to the social assessment of technology, which is an interdisciplinary field developed the United States and Europe (Hetman 1978). This field is preconditioned by political, economic, sociological, psychological, sociocultural, and ethical aspects of the development of technologies, as well as the implementation of specific technical projects. The problems of applied ethics, although directly adjacent to the discourse of this field, can be more directly attributed to the philosophy of technology. Defined more specifically in disciplinary terms, applied ethics analyzes the problems that arise during the development and implementation of techniques and technologies in terms of ethical dilemmas, offering solutions to the problems created by the technologies. These proposals can then be utilized by authorities in interdisciplinary fields to assess the social acceptability of certain technologies, moderated by constraints they derive from other related fields.

[1] Vitaly Pronskikh, PhD (Physics), PhD (Philosophy), Fermi National Accelerator Laboratory, PO Box 500, Kirk Rd & Pine Str., Batavia, 60510 IL, USA; vpronskikh@gmail.com

The problem of choosing which variant of the NFC (or which combination of variants) to implement is one of the most important problems all countries of the world face when using nuclear energy (International Atomic Energy Agency [IAEA] 2019; Cho 2020). In total, there are thirty-one countries in the world that produce energy using nuclear power plants (although that number is falling; Germany, Belgium, and Switzerland are decommissioning such power plants), but their approaches to the implementation of NFCs differ significantly. The balance between the benefits they acquire from the sale of electricity and the costs, which include (in addition to direct economic considerations) the costs of raw materials, production, waste disposal, specialist training, and the safety risks of nuclear energy at all stages of its use from ore mining to nuclear waste disposal, plays a determinative role in such a choice.

Despite the abundant complexity, risk, and even pessimism often associated with nuclear power in a number of European countries, its fate is not yet completely sealed. In particular, the Joint Research Centre's Report for EU taxonomy for sustainable activities (EU 2022) optimistically states, "The fatality rate caused by severe accidents is for nuclear energy comparable to, and for the Gen III NPPs lower than, that of any other electricity production technologies and that the maximum consequences of a single event are rather high but still comparable with some other electricity production technologies." Note, however, that these conclusions relate to slow neutron reactors used in the open NFC, which will be discussed more thoroughly in the following section. Although some countries that were planning to decommission nuclear power plants have suspended these activities, there is not yet sufficient evidence to suggest that the construction of new power plants in countries that have expressed such intentions in the past will be resumed or completed in the near future.

Several interventions have aimed to reduce the proliferation risks of weapons-grade uranium by repurposing it for the nuclear energy production. One of the most successful efforts of this kind was the "Megatons to Megawatts" program that began in 1993 and ended in 2013. Over these two decades, Russian Highly-Enriched Uranium (HEU) was down-blended into Blended Low-Enriched Uranium (BLEU) fuel in the US. Almost 500 metric tons of HEU were processed during the program, significantly reducing the amount of weapons-grade uranium in the world and, therefore, potential risks of its military use (Slater-Thompson and Bonnar 2013).

The consequences of a particular NFC are not always easy to evaluate economically. Various NFC options (nominally open and closed) carry different risks, so deciding which risks are more acceptable is explicitly and implicitly based on ethical principles. Moreover, these risks pose additional ethical concerns because they are unequally distributed among different regions and generations. Countries also consider ethical concerns arising in the context of uranium resource depletion and nuclear waste management as well as inevitable risks for present and future generations. These concerns include inter- and intra-generational justice—Stephen Gardiner's perfect moral storm (Gardiner 2003)—and the problem of the identity of future generations—Derek Parfit's nonidentity problem (Parfit 1983). Our objective was to analyze the ethical grounds for certain preferred options for the implementation of NFCs in the world based on a comparative analysis of safety risks at various stages of an NFC.

## 2. NFC Stages and Safety Risks

The depletion of raw materials for energy production is an obvious problem for the Earth's future population. At the same time, scholars often note that for certain nuclear energy production technologies (such as Fast Reactors (FRs)), reserves of fuels such as natural uranium-238 can last for a thousand years. In the case of more traditional thermal neutron reactors (like Light Water Reactors (LWRs) and other reactors using Low-Enriched Uranium (LEU) fuel), we may exhaust the reserves of suitable raw materials much earlier. However, there are several serious problems that complicate this comparison. First, FRs are more complex and difficult to operate, and they pose significantly more risks to the populations where they are sited. Second, governments could use such reactors to produce weapons-grade plutonium-239. For these reasons, most countries that have ever operated FRs (the United States, France, Germany, and Japan) have decommissioned them. Currently, some FRs are operating in Russia, and India and China are considering the construction of two more. The advantages of FRs include their ability to destroy (transmute) some types of nuclear waste by using them as fuel. However, these advantages must be weighed against the increased radiological and security risks for the local population. It is noteworthy that virtually all plutonium formed in spent fuel from conventional LWRs and other reactors using LEU fuel (unlike FRs) becomes unsuitable for weapons use after typical fuel cycle durations. This is because, in addition to Pu-239, which is required in large amounts for military purposes, the spent fuel contains a significant fraction of Pu-240, which inhibits the relevant nuclear chain reactions. To avoid this disruption, nuclear proliferators may attempt to cycle the fuel in a reactor during abnormally short periods and then remove it from the reactor core for reprocessing. However, such a practice deviates from the standard commercial power plant operation protocol, and international monitoring authorities like the IAEA could easily notice and provide sufficient grounds for curtailment. Thus, there exist rather profound philosophical grounds underlying most countries' refusal to introduce riskier, albeit more resource-saving, technologies at the present time.

NFCs can be classified into two broad variants: open and closed. We will consider schematically, for the purposes of our discussion, their main stages (see, for example, IAEA 2019).

**2.1 Stages of an Open NFC**

An open NFC includes the following main stages:

1. uranium mining and production of uranium oxide $U_3O_8$ (yellowcake),
2. conversion of $U_3O_8$ to $UF_6$,
3. enrichment of uranium (by isotope U-235),
4. production of fuel,
5. energy production (for example, in a thermal reactor),
6. interim temporary storage,
7. preparation of nuclear waste for storage (low-level waste bypasses stage 6), and
8. long-term storage of nuclear wastes.

**2.2 Stages of a Closed NFC**

Steps 1–5 are similar to those for an open NFC. These steps then follow:

6. separation and recycling,
7. sending the extracted uranium to stage 2,

8. using the extracted plutonium to manufacture mixed oxide (MOX) fuel (together with uranium coming from stage 2), which is then sent to an FR or LWR,
9. preparation of nuclear wastes for storage, and their long-term storage.

**3. Risks for Different NFCs**

We considered the two most common types of NFC risks: 1) Environmental and public safety risks arise at all stages of the extraction and manufacture of nuclear fuel, the operation of a nuclear power plant, its decommissioning, and the storage and disposal of spent fuel. These risks are associated with possible accidents that lead to environmental pollution, as well as human exposure while handling nuclear fuel and nuclear wastes; and 2) safety risks associated with nuclear material theft and/or sabotage leading to the dispersal of radioactive materials that could occur at any of the above stages of material handling. Behnam Taebi's (2012) detailed analysis shows that in relation to the risks of the first type, a closed NFC with light-water FRs, breeders, and transmuters has an advantage over an open cycle in only one aspect—fuel extraction (stage 1), which is highly risky in the second case and low in the first. With regard to fuel transportation, commissioning, and decommissioning, the open NFC turns out to be much safer. As for the risks of the second type, the FR-transmuter, with respect to its operation in a closed cycle, provides a similar level of safety to the open one, despite a FR-breeder (or Light Water Reactor – Fast Reactor, LWR-FR) carrying significantly more risks. The technological implementation of a closed NFC with any FR is much more complicated. Thus, it is possible to summarize the observations of Behnam Taebi and Andrew Kadak (2010) as follows: a closed cycle, although more economical, is generally much riskier than an open one.

## 4. Ethical Arguments and Their Relation to the Problem of the Choice of NFC

By expending uranium reserves in the present, humanity depletes them for the future, and the nuclear waste storage facilities required due to uranium use pose an obvious danger to both present and future generations. However, to understand exactly what ethicists say is or is not permissible and why, we appealed to ethical theories. Among the ethical theories that find the most frequent application in the assessment of nuclear safety risks, one can include, for example, the utilitarianism of Jeremy Bentham (1748–1832), which demands an evaluation of the possible total benefits and risks from the introduction of any technology for a group of people (population) and considered morally justified when those whose comparative benefits outweigh the risks. In a number of other issues it may possible to rely on the doctrine of deontology (proposed by Immanuel Kant) which judges the rightness or wrongness of an action against a series of universal rules regardless of its consequences. The precise establishment of universal norms (for example, radiation safety) for each individual resident of the country must certainly be fulfilled, because it is a moral duty, nonetheless, in the case of NFCs, the establishment of universal norms for all countries is indeed still unattainable. However, even relying on utilitarianism and minimizing risks in comparison with benefits, we face dilemmas, because the risks are not the same for different groups of the population, and the benefits received by the current generation create a burden on the shoulders of future generations. To solve the question of how to better distribute risks and benefits in this case, we must turn to the theory of justice.

According to the IAEA (2019), nuclear power plants produce about 10 percent of all electricity consumed in the world. At the same time, there are over 6 million cubic meters of nuclear waste in storage facilities. Radiotoxicity of transuranic isotopes from spent fuel (in the absence of their extraction and processing), although it will decrease over time, will still be traceable for ten thousand years. The largest underground storage of such radioactive waste in the United States is located near Carlsbad, New Mexico and is situated at a depth of six hundred meters in rooms created in a salt mine. However, exactly what part of spent nuclear fuel is waste depends on whether some of its components can be returned to power plants as fuel after processing. The same can be said for plutonium-239 from weapons stocks; it, too, can be processed into fuel for some types of nuclear power plants (for example, by mixing plutonium oxides with uranium oxides into MOX fuel). The problem is that such secondary fuel obtained from waste or weapons stocks can be used in either FRs, which are riskier to operate, or in some conventional thermal (or slow) neutron spectrum reactors. One of the risks is that in some modes of their operation and designs, they can themselves become so-called breeders, producing new plutonium-239 with weapons application and thereby creating even greater proliferation risks.

An FR is not the only reactor type that can be used to burn MOX fuel. This type of fuel can also be used in Light-Water Boiling Water Reactors (BWRs), similar to those that were part of Fukushima Daiichi and were damaged in the Great Tōhoku Earthquake and tsunami. Another type of reactor that can be fueled with MOX is a Pressurized Water Reactor (PWR). The development of a fleet of LWRs capable of operating on MOX fuel consisting of down-blended plutonium pits from weapons stockpiles was part of the US nuclear power program. In the US, a MOX fuel production facility at the Savannah River Site (SRS) was proposed, and the Tennessee Valley Authority analyzed MOX fuel performance at Browns Ferry BWRs (O'Grady 2011). In 2018, the National Nuclear Security Administration terminated the US MOX facility contract, mainly for environmental and safety reasons (see, for example, WNN 2018). Hence, regarding MOX fuels, environmental and proliferation risks are at stake, which affects both FRs and LWRs equally. These risks are not specific to the open NFC.

Choosing an NFC establishes not only a specific set of processes from uranium mining to waste disposal but also the kind and amount of waste produced during the operation of a nuclear power plant. Two ethical problems are central to this decision: resource conservation and safety of use.

**5.   The Problem of Resource Conservation**

Discussion about the problem of resource conversation can be based on a parallel discussion about climate change problems, which Taebi (2012) applies to the analysis of nuclear technologies. John Rawls ([1971] 1999) initially raised the question of so-called intergenerational justice (a kind of distributive justice), convincingly arguing that any generation of the world's population needs to treat their descendants fairly, without depriving them of the same benefits (such as natural resources) enjoyed by the current generation. Subsequently, Gardiner (2003) extended this concept to climate change, pointing out that the burning of hydrocarbon fuels benefits the current generation but leaves future generations with the postponed, intractable problems of global warming. Even in a case when greater benefits in the present bring smaller problems in the future, this, according to Gardiner (2003), represents a moral problem.

By analogy with the even more famous prisoner's dilemma—in which two prisoners each act against their own self-interest by not betraying the other in exchange for a reduced sentence—Gardiner formulated two statements. Each statement sounds logical on its own, but there is a substantial inherent contradiction between them. The first statement asserts that it seems logical for each generation to preserve and transfer natural resources to future generations because such behavior maximizes the benefits for each of them. The second asserts that from an individual point of view, it is logical for each generation to maximize the use of available resources for its own purposes. Self-centered behavior by one generation leads to the loss of resources for subsequent generations.

Philosophers in search of an ethically acceptable solution to the intergenerational problem often turn to Brian Barry's (1989) two ethical principles: the principle of responsibility and the principle of vital interests. According to the principle of responsibility, if a person's fault in the occurrence of some problem is not proven, then they deserve compensation for damage from this problem. According to the principle of vital interests, a person's position in space and time (for example, belonging to a particular generation) cannot serve as a basis for violating their interests that are essential to life and survival. Within this specific context, it is imperative not to assume that the mere existence of modern technology inherently encompasses the essential concern of a generation. The preservation of a future generation's quality of life should not be equated with ensuring their survival, as such a perspective oversimplifies the matter. Nevertheless, as we delve into the subsequent discussion on the nonidentity problem, we realize that this philosophical inquiry equally applies to both these notions, blurring the boundaries between them and emphasizing their inherent similarity.

If the conditions of the principles mentioned above are met and future generations receive intact (or sufficient) natural resource reserves and a preserved environment from previous generations, then no compensation will be needed. However, if these conditions are not met, it is unclear what compensation past generations could offer or how future generations could even collect. To put it differently, while the principles of responsibility and vital interests are capable of guiding ethical behavior, they fall short in terms of meting out justice for past generations.

A noteworthy aspect of Barry's egalitarian justice is that it requires an equal distribution of benefits only in the case of initially equal opportunities for individuals (or generations). Thus, applying this theory to the discussion of intergenerational justice, it becomes necessary to clarify that the main idea is to demand equal opportunities for different generations. This allows for a broader interpretation than just the equality of benefits. Moreover, in addition to the preservation of natural resources, technological resources are also conserved and passed down to future generations, offering direct advantages and prospects. These prospects may even encompass the opportunity for future generations to address complex technological challenges of the present, such as the management of spent nuclear fuel. In this regard, the preservation of technologies like FRs)holds instrumental value, as they serve as means to an end (intrinsic value, on the other hand, pertains to the preservation of nature for its own sake, independent of humanity's interests). One possible solution would be to preserve and develop computer models (simulations) of technologies that are being phased out (Pronskikh 2022).

Extending the discussion of distributive justice to nuclear power, Taebi (2012) and Taebi and Kadak (2010) pointed out that a similar problem arises because 1) uranium is a nonrenewable resource, and its extraction by the current generation impoverishes its reserves for future generations; and 2) the accumulation of nuclear waste, a matter of convenience for the current generation, also creates environmental problems for future generations. In view of this, Taebi (2012) stated that our obligations to future generations are to limit the extraction and use of uranium as well as to ensure the safe storage of waste. From the vantage point of the pure intergenerational problem, Taebi's (2012) proposals to reduce uranium production and switch to the recycling of irradiated fuel sound reasonable (i.e., to prefer the closed NFC to the open one).

However, there are other important philosophical arguments that we believe contradict Taebi's conclusions. One such argument is based on attempts to resolve Parfit's (1983) nonidentity problem. The fates of individuals in Gregory Kavka's "The Paradox of Future Individuals" (1982), which Melinda A. Roberts ([2009] 2019) discusses in more detail, illustrate some of the main features associated with the non-identity problem. The conclusions can be generalized from individuals to communities and generations. According to the example of Kavka (1982), a fictitious poor couple enters into a contract with a rich man to birth a child and hand it over to him as a slave. If the couple refrains from completing such an immoral contract, then they can make a choice: either give birth to a child (without the contract) or not (also without a contract). The first alternative seems to be irreproachable; however, as the proponents of Parfit's approach will note, the child born in this case would be a different person, not a slave, so he or she would not be identical (will be non-identical) to a slave child born under the contract. Assuming axiomatically that the second alternative (not giving birth to a child) is obviously worse than all the others (since Parfit accepts that existence is, in principle, always better than non-existence), one can arrive at the conclusion that giving birth to a child under a contract is an acceptable alternative to not giving birth to a child. Bracketing the ethical assessment of a particular life as good or bad, Parfit's solution to the problem suggests that in situations that involve life-changing choices, the only actual alternative to existence in relatively bad conditions is non-existence (either due to not being born or due to the birth of someone with a different identity, which would be equivalent to non-birth).

This problem can be linked to the discussion of nuclear energy by considering Parfit's example of planetary resource depletion. His position was that neither in the development of technology nor in other human actions can there be anything inherently good or bad, rather only good or bad for certain individuals or groups. Thus, on the one hand, if the current generation depletes natural resource reserves (such as uranium) to maximize their own benefit, then they would greatly disadvantage future generations by requiring them to extract resources with great difficulty or invent fundamentally different energy technologies to maintain a similar standard of living. Such a perspective requires maximizing resource conservation.

On the other hand, if humanity conserves resources by radically changing its lifestyle (e.g., by stopping mining), this will also radically affect future generations. In this changed world, people of the future would be born different, at a different time and with different personal and physiological qualities—they will be different people from those who would have been born if everything remained the same. Other people of the future, if they exist, will also create a culture (norms, values, languages) different from what would have arisen without such changes in the past (similar to Ray Bradbury's butterfly effect). Now, assume that to prevent the exhaustion of energy resources and mitigate climate change, a country stops using carbon-intensive methods of energy consumption such as rail, auto, and air transport, and even closes its borders with other countries. In this scenario, communications between regions and countries would virtually stop, and disunity and regionalism would increase. The descendants would have a different origin and belong to cultures other than if the transport had remained the same. Thus, a drastic restriction of the extraction of minerals, in the absence of alternative equivalent energy sources, will mean that future generations will be made of people fundamentally different from those who would have appeared with the continuation of their extraction. It is for this reason that Parfit's argument is called the nonidentity problem. Parfit makes a strong theoretical argument which many ethicists explicitly or implicitly endorse with only few reservations regarding particular cases and examples. This argument challenges the validity of Taebi's (2012) proposal that our obligations to our descendants consist in limiting uranium mining, because in the event of any pivotal restrictions, the future will change.

## 6. The Safety Problem of Using NFCs

The second ethical dimension of NFCs considered in this analysis is the problem of handling nuclear wastes. As discussed above, in an open NFC, spent nuclear fuel (SNF) is secured for temporary and then for long-term storage. In the case of a closed NFC, uranium and plutonium are removed from the SNF and are returned to the fuel cycle (for example, in the form of MOX fuel for FRs or BWRs). Minor actinides (MA), including neptunium-237 and plutonium-239 can be isolated from the SNF, and then the remaining nuclear wastes are sent for long-term storage. The remaining nuclear wastes are mainly constituted by long-lived fission fragments (LLFF) including such large-yield elements as technetium-99, zirconium-93, cesium-135, iodine-129. MA, after afterburning in the neutron fields of reactors, also turn into LLFF, which are received for long-term storage. The radiotoxic lifetimes of the SNF components (compared to the initial uranium ore) are as follows: ~200,000 years for the initial SNF (discharged from the reactor), ~10,000 years for MA and LLFF and curium (i.e. after the removal of uranium and plutonium), ~500–1000 years after the removal of MA (LLFF and curium), and ~200-500 years for LLFF. Thus, by removing and afterburning uranium, plutonium, MA, curium, in a FR, it is possible to make nuclear wastes dangerous only for 200-500 years. However, we should note that with existing technologies it is impossible to completely recycle MA in FR, and additional nuclear waste is also generated during the manufacture of fuel from recycled plutonium. That being said, the risks to the safety of NFCs in the case of nuclear fuel processing (closed NFC) are much higher than in the case of an open NFC, where even in near-surface storage facilities safety can be ensured for 20,000 years.

In a report of the National Academy of Public Administration for the US Department of Energy (NAPA 1997), the authors indicated that when comparing nuclear technology risks between 2-4 generations and human societies 500-1000 years apart, preference should be given for the interests of populations living closest to present day. Since the uncertainties with distant generations are so great, it is impossible to reliably predict their nature, their technological capabilities, or how certain risks may affect them. The authors do assume that far-future generations will have probably figured out how to optimally deal with radioactive wastes with the aid of advanced technologies.

Similar standards are adopted in European countries. For example, in the report of the Swedish Advisory Committee on Nuclear Waste Management to the Swedish Ministry of Environment and Energy (see Taebi 2012), future generations are conditionally divided into three periods: 1) distanced by 150 years or less from the present, 2) distanced by 150–300 years from the present, and 3) distanced by more than 300 years from the present. The duration of a generation is thirty years, and 150 years is the duration of five generations. This timespan corresponds with the human capacity to emotionally empathize with descendants and have coherent assumptions about their character—that is, up to the grandchildren of one generation's own grandchildren. At the same time, 300 years is assumed to be a characteristic period of the existence of a nation, during which they are able to preserve their unique features and a person is able to sympathize with descendants as sharing the same national identity (although this is a rather speculative assumption); after a further 300 years, it is impossible to predict the identity of the population. For this reason, the Swedish model demands maximum fairness in the allocation of resources for the first group, the average for the second, and the minimum for the third. The security requirement is also distributed in a similar way. Thus, not all generations are considered equal in terms of ensuring their safety. From this viewpoint, the safety of distant generations is less of a priority than that of nearer ones.

## 7. An Ethical Analysis of NFCs

We believe that the previously mentioned standards in both the USA and Sweden are in line with the problem of Parfittean nonidentity. If the change in the existential conditions of the current generation is minimal, then it becomes possible to predict the traits of descendants for a longer period. Based on historical examples, the way of life in traditional societies and their identity often do not change over long periods of time. Only abrupt, drastic changes in living conditions, as a rule, were accompanied by identity metamorphoses. If we assume that it is difficult to predict the identity of the population of a particular region beyond the 300-year period, then currently, because of the rapid processes of migration in the world, it becomes difficult to predict it even for a period of less than 150 years. From this point of view, preserving the level of extraction and use of minerals (and the standard of living) of the current generation and reducing risks for them becomes an increasingly attractive and ethically acceptable strategy. This will be satisfied by a reduction in SNF reprocessing (and the use of less safe FR) and a preference for long-term storage. This seems to us a plausible explanation why an open NFC is preferred in many developed nuclear energy producing countries.

Recycling and closure of the cycle means transferring all the increased risks from the implementation of such a cycle to the shoulders of the current and near generations to protect generations that are more than 200-500 years away from the current one (determined by the time of reduction to a safe level of radiotoxicity of LLFF with complete processing). However, the timescale of even 200 years comprises the period of identity preservation by local communities (in the Swedish model) and, against the background of the social processes discussed above, indeed overestimates that period. Nevertheless, the implementation of complete SNF reprocessing (complete closure of the cycle) will create higher risks for the current and near generations implementing such a cycle. The key issue is the safety of the closed NFC at all stages of the cycle, which currently cannot be guaranteed. Radiation accidents caused by using a closed NFC (during recycling, FR operation, and decommissioning) could be so significant that they will change the way of life in large areas more radically than even the Chernobyl accident. That explains why many agree that the first priority is to protect the current generation.

Some countries plan an alternative to the choice of NFC—the complete abandonment of nuclear energy. Other countries are implementing intermediate options to minimize risks; for example, by purchasing uranium raw materials extracted in other countries, exporting nuclear waste to other countries for processing, or implementing the build, own, operate (BOO) scheme, in which another country builds nuclear power plants, brings nuclear fuel, maintains the nuclear power plant, and then takes the SNF to itself for processing and disposal. Some countries prefer to buy electricity from other countries without developing their own nuclear power. Obviously, such schemes reflect the global division of labor, but there are critical assessments of this situation. Mike Hannis and Kate Rawles (2013) propose three criteria under which the compensated transfer from one country to others (recipients) of nuclear wastes for processing and disposal is ethically acceptable and cannot be considered a bribe to the elites of these countries, governments of which are expected to use the compensation to mitigate risks. The criteria include 1) choosing a safe place for processing and disposal of nuclear weapons regardless of compensation, 2) taking into account the interests and consent of the population of the recipient countries of nuclear wastes and their descendants, and 3) discussing compensation without the direct participation of nuclear energy companies (to ensure disinterest). The global trend is that the more economically prosperous countries of Europe either plan on abandoning nuclear energy or minimizing risks by shifting them to other countries in one way or another; economic and technologically developed countries (for example, the US) use an open NFC and do not utilize FR. All this is also consistent with these countries' prioritization of the current generation's safety. We believe these ideas are largely consistent with the ethical principles analyzed above.

Two innovative FR varieties that are still under development are Small Modular Reactors (SMRs) and microreactors (Patel 2020). Developers are proposing the use of High-Assay Low-Enriched Uranium (HALEU) fuel for these, which can be enriched up to ~19.75% of U-235. Their output power can start from 10 MW; such reactors can operate continuously for up to twenty years without needing to load new fuel. For example, the design of TRISO-type of HALEU fuel elements for such reactors "consists of a very small kernel of uranium coated with a variety of silicon- and carbon-based materials, and is designed to withstand extreme heat with low proliferation concerns and environmental risks" (Patel 2020). The development of SMRs and microreactors may require rethinking NFC risks. On the one hand, additional benefits from such reactors can be envisaged (after they are successfully built, tested, and demonstrated), as they might create additional opportunities for future societies in the near future and, probably, the long term. On the other hand, their SNF will probably require a longer cooling-off time before transferring to dry casks, that is, long-term storage in wet storage nuclear facilities. Also, in virtue of their small size, relative ease of operation, and potentially wide distribution, proliferation risks may not be associated so much with unauthorized access to SMRs' and microreactors' SNF elements as the reactors themselves. In addition, the risk of rare but significant accidents at traditional large reactors may be supplemented by the risks of multiple local accidents.

A rather challenging question is whether it is possible and necessary to re-tool existing closed NFC facilities (for example, those employing FRs) in the countries that are building or operating them. The examples discussed here are Japan, France, and Russia. On the one hand, from a normative standpoint, the ethical framework developed in this work is sufficiently universal to be applicable to existing facilities to the same extent as future ones. With respect to the distant generations, there should be no difference between the damaging effects of the existing reactors and the ones built in the near future. On the other hand, regarding how practical it is to expect policy changes in certain countries, it must be recognized that the difference mainly lies in aspects of national culture. In particular, Taebi and Kadak (2010) introduce a distinction between intrinsic and instrumental values, which we associate with the well-known division into internal and external. Intrinsic (or internal) values are related to the theories that ascribe to nature (or environment) its own value, not associated with benefit to anyone else. Instrumental (or external) values are directly related to the benefits of certain actions for a person, since such theories consider nature only from a utilitarian and pragmatic point of view as a source of benefits. For example, Taebi and Kadak (2010) themselves adhere to an externalist point of view, while IAEA sticks to an internalist, discerning the safety of technologies for humans and the environment. If one turns to the cultural contexts of the aforementioned countries, one can assume that France would perhaps take a moderately externalist position, perceiving human modification of the environment as an inevitable phenomenon (Latour 1988) although undertaking substantial efforts to mitigate harm. Russia would likely take the most radical externalist position, since the history of the Soviet regime has essentially shaped society's attitude toward nature as, on the one hand, a workshop, and, on the other hand, due to the vastness of undeveloped territories, as a resource that is always in abundance (Golden 1982). Japan, with a centuries-old history of predominating animist beliefs, is likely to view nature as endowed with divine attributes (Toshimaro 2004). Assuming such reasoning to be plausible, one could expect that these countries would consider re-tooling their existing installations from closed to open NFC in the following order of decreasing confidence: Japan, France, and Russia. It should, however, be noted that drawing conclusions based on cultural traditions can be onerous, especially since some confounding factors are difficult to fully take into account, such as the financial and political situation. Such reasoning would rest on rather shaky ground. Thomas P. Hughes' concept of technological momentum (Hughes 2000) enables another argument in favor of repurposing existing closed NFC to open. According to Hughes, a newly developed technology allows for deliberate control from the society. However, after it matures, it becomes deeply embedded in the society that created it to such an extent that its previously inherent lack of determinism and controllability vanish by virtue of its technological momentum. Since the technologies of closed NFC are currently at the stage of development, testing, and pilot projects, one can assume that they have not yet gained sufficient technological momentum and thus can be radically changed or stopped from implementation. If such technologies become basic in certain countries (or their associations), then stopping them, if necessary, would become extremely difficult, if not impossible.

   From this point of view, we find the BOO scheme interesting for analysis. According to this scheme, a country builds nuclear power plants in the territory of other countries, and by retaining ownership, ensures their operation and decommissioning upon completion of their contract (see also Pronskikh 2020). This approach saves the contracting countries from needing to train their own personnel to manage the reactor and from having to purchase fresh fuel and dispose of the SNF (because it is delivered from the supplier country and returned after it is spent). In other words, according to such a scheme, the supplier country uses its resources to create and implement electricity production on the territory of another country. The fundamental difference between such a model and the traditional trade in energy resources (e.g., oil and natural gas) is that the supplier country must dispose of and bury on its own territory nuclear waste produced abroad. For this reason, we believe governments should carry out an ethical examination of this scheme, accounting for the conclusions and recommendations made by Hannis and Rawles (2013). They should ensure the disinterest of the parties involved in the compensation discussion, as well as considering the interests of the population's descendants (even thinking of only near ones according to the Swedish model) living in the territory where they will build storage facilities for imported radioactive waste. One could argue that such a scheme poses risks to both present and future generations, although it has not yet reached its technological momentum to enable more precise predictions.

## 8. Conclusion

We examined the traits of open and closed NFC, analyzed their risks, and discussed studies showing that the implementation of a closed NFC leads to significantly higher risks than an open one. The problem of the choice of NFC is reduced to the resolution of two related problems, resource conservation and safety of use (including the problem of handling nuclear wastes). We analyzed ethical arguments and theories that underlie approaches to solving these problems, issues of intergenerational and intragenerational distributive justice and responsibility to future generations, the problem of nonidentity, and we discussed the features of NFC based on these considerations. We showed that the global trends of preference for an open NFC over a closed one, the refusal of countries to engage in SNF processing, their shipping of SNF to other countries, or even their complete rejection of nuclear energy are grounded in ethical considerations about the priority of the current population's interests and safety compared to the interests of distant generations with a different identity. We suggest that open debate between supporters of the Swedish model and supporters of the well-being of far-future generations seems to be the demand of the time. We argued that reducing uranium use is not a feasible solution to the problem of nuclear wastes, because it can entail the nonidentity problem of future generations even in a relatively short term. Although the closed NFC possesses certain advantages in terms of more efficient use of resources, to make it ethically acceptable for most countries would require more developed technologies for the recycling and reprocessing of SNF that are equal or superior to open NFC technologies in terms of safety, security, and ecological friendliness.

**Acknowledgements**

Fermi National Accelerator Laboratory is operated by the Fermi Research Alliance, LLC under Contract No. DE-AC02-07CH11359 with the U.S. Department of Energy, Office of Science, Office of High Energy Physics. The author is indebted to the Editor, Taylor Loy, and two anonymous reviewers for their careful reading of the manuscript and many insightful comments and suggestions.